\documentclass[aps,prl,twocolumn]{revtex4}   
\newcommand {\be}{\begin{equation}}
\newcommand {\ee}{\end{equation}}
\newcommand {\bey}{\begin{eqnarray}}
\newcommand {\eey}{\end{eqnarray}}
\usepackage{epsfig}
\usepackage{amsfonts}
\usepackage{mathbbol}

\begin{document}

\title{Quantum decoherence reduction by increasing the thermal bath temperature}
\author{A. Montina, F.T. Arecchi}
\affiliation{Dipartimento di Fisica, Universit\`a di Firenze,
Via Sansone 1, 50019 Sesto Fiorentino (FI), Italy}

\date{\today}

\begin{abstract}
The well-known increase of the decoherence rate with the temperature,
for a quantum system coupled to a linear thermal bath, holds no longer 
for a different bath dynamics. This is shown by means of a simple 
classical non-linear bath, as well as a quantum spin-boson model.
The anomalous effect is due to the temperature dependence of the bath 
spectral profile. The decoherence reduction via the temperature 
increase can be relevant for the design of quantum computers.
\end{abstract}
\maketitle
The rapid development of nanotechnology has opened
the possibility of assembling materials and devices at the
atomic scale,
such as nanomaterials and nano electro-mechanical systems 
(NEMS)~\cite{Blencowe}. 
Because of quantum mechanical and statistical effects,
the properties of these systems can be considerably different from
their macroscopic counterpart.  Electrons in nanofabricated quantum 
dots as well as nanomechanical devices are possible candidates for 
quantum bits at nanometer scale, as suggested by recent 
experiments~\cite{garcia}. Quantum computers could take 
advantage of quantum superpositions in order to perform highly 
efficient parallel calculations. The main difficulty in the 
implementation is represented by decoherence, due to the 
coupling with the environment~\cite{decohe}. This effect increases 
with the number of quantum bits and is devastating 
even with a small thermal bath coupling. Thus, in order to
reduce the decoherence of quantum superpositions, 
a sufficiently small thermal disturbance and 
environment coupling are required. The thermal disturbance is 
expected to be reduced by decreasing
the bath temperature. This is true when the quantum system is 
coupled with a high number of degrees of freedom and the bath can 
be described as a linear noise source. On the contrary, at the
nanometer scale, the quantum system can be directly coupled to
few modes, whose power spectrum does not necessarily 
grow with temperature. Thus, it is possible to have a reversed
effect, i.e., the reduction
of decoherence by increasing the thermal bath temperature.

A similar counterintuitive behavior occurs for example in
thermal rectifiers~\cite{casati}, where the heat flux can be suppressed
by increasing the temperature difference of the junctions.
The rectifier consists of two or three mutually coupled 
layers, whose phononic bands depend on the temperature via
non-linear effects. As the ends of the chain are
coupled with two heat baths at different temperatures $T_1$ and $T_2$, 
the overlaps among the phononic bands in each layer become functions
of $T_1$ and $T_2$ and are not identical when the thermal baths
are exchanged. As a consequence, the thermal flux can change
when the rectifier is reversed.  

In this paper, we show that the decoherence rate of a quantum system 
coupled to a non-linear thermal bath can decrease with the temperature. 
As in thermal rectifiers, this effect is due to the non-linearity, 
which makes the shape of the bath spectrum dependent on the 
temperature.
For our purpose, we consider a classical non-linear bath model, 
consisting of a one-dimensional particle in a double-well potential, 
strongly coupled to a linear bath. Furthermore, we also consider a full 
quantum mechanical bath model, consisting of a boson bath coupled to a two-state 
system, which is the low temperature limit of a double-well system under 
suitable conditions~\cite{leggett}. A probe quantum system is coupled to the 
non-linear bath and its decoherence rate is evaluated. 

The equation of motion of a one-dimensional particle coupled to a white noise thermal 
bath is $m\ddot x=-V'(x)-m \bar\gamma\dot x+\sqrt{2 D}\eta(t)$,
where $m$, $V(x)$, $\bar\gamma$ and $D$ are the particle mass, the external 
potential, the dissipation and diffusion coefficients, respectively. The 
function $\eta(t)$ is a Gaussian noise with $\langle\eta(t)\eta(t')\rangle=\delta(t-t')$.
The fluctuation-dissipation theorem establishes that $D=\bar\gamma m k_b T$, 
$k_b$  and $T$ being the Boltzmann constant and the thermal bath
temperature, respectively. We consider the double-well potential
$V(x)=-a x^2/2+b x^4/4$.
By means of the rescaling $x\rightarrow (a/b)^{1/2} x$, $t\rightarrow (m/a)^{1/2}t$
and $T\rightarrow \frac{a^2}{b k_b} T$, the equation of motion becomes
\be\label{stocha_norm}
\ddot x=x-x^3-\gamma_1\dot x+\sqrt{2\gamma_1 T}\eta(t),
\ee
where $\gamma_1\equiv\bar\gamma (m/a)^{1/2}$. The independent parameters are the
rescaled dissipation coefficient and temperature. 

\begin{figure} [h!]
\epsfig{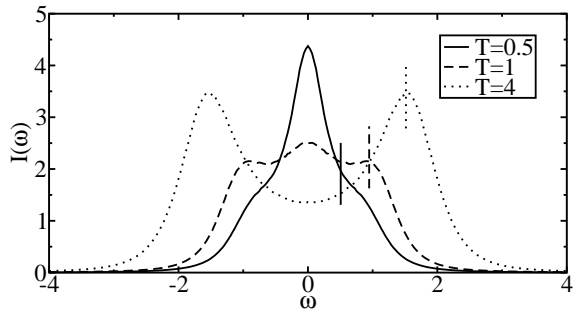}
\caption{Power spectrum of $x(t)$ for some values of $T$. The vertical lines are
placed at $\pi R$ for each spectrum.}
\label{fig1}
\end{figure}

\begin{figure} [h!]
\epsfig{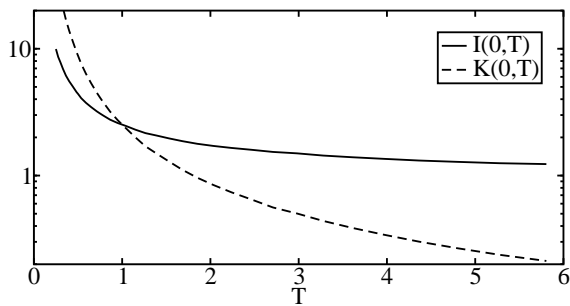}
\caption{$I(0,T)$ and $K(0,T)$ as functions of the temperature, in logarithmic scale.}
\label{fig2}
\end{figure}

\begin{figure} [h!]
\epsfig{figure=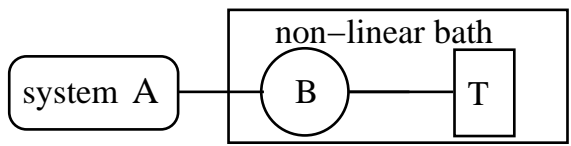,width=7.5cm}
\caption{Schematic representation of the non-linear thermal bath, coupled to
the quantum system $A$. $B$ is the anharmonic oscillator.}
\label{fig3}
\end{figure}

At finite temperature, the particle has a finite probability to jump from a well
to the other one. The hopping rate is~\cite{sto_res}
$R\equiv(\sqrt{2}\pi\gamma_1)^{-1}\exp[-(4 \gamma_1 T)^{-1}]$ (Kramers rate).
In the limit of very low temperature, the hopping rate becomes very small 
and the particle remains in one well for a large time. As a consequence,
the spectrum of $x(t)$ has a high zero-frequency 
peak. As the temperature is increased, the hopping rate between the two wells
grows, the central peak of the spectrum reduces and two smooth peaks around the
frequencies $\pm \pi R$ are generated. This behavior is shown in Fig.~\ref{fig1}, where
we report the spectrum 
$I(\omega,T)=2 \text{Re}\int_{0}^\infty \langle x(0) x(t)\rangle_{T} e^{i\omega t} dt$,
for $\gamma_1=0.4$ and some values of the temperature. $I(\omega,T)$ has been evaluated
numerically over a time interval of $200$, in the non dimensional units, and $50000$ ensemble
realizations. We have verified that the time interval is sufficiently large to 
ensure a negligible time cut-off of the integral.
In Fig.~\ref{fig2}, we plot
$I(0,T)$ (solid line) and $I(0,T)/T\equiv K(0,T)$ (dashed line) as functions of $T$. 
The decreasing behavior of $K(0,T)$ suggests a simple model of
thermal rectifier, as will be discussed in an extended paper. Here
we will use the decreasing behavior of $I(0,T)$ to show the
anomalous behavior of decoherence induced by the non-linearity.
We consider a two-state quantum system ($A$) coupled to the double-well
oscillator ($B$), as sketched in Fig.~\ref{fig3}.
The overall system $B$-(linear bath) is the non-linear bath seen by $A$.
The linear bath and the anharmonicity of $B$
guarantee that the overall system is a reservoir and a non-linear system,
respectively.

This classical bath model is suitable for a sufficiently high thermal
bath temperature $T$. A full quantum approach will be introduced later on.
We consider an interaction Hamiltonian linear in $x$ and assume that the quantum 
state $|\psi\rangle\equiv\psi_{-1}|-1\rangle+\psi_1|1\rangle$ satisfies the 
Schr\"odinger equation
$i\frac{\partial|\psi\rangle}{\partial t}=\left[(\Omega/2)\hat\sigma^1_A+
\epsilon\hat\sigma^3_A x(t)\right]|\psi\rangle$,
where $\hat\sigma^k_A$ are the Pauli matrices in the orthonormal basis
$\{|-1\rangle, |1\rangle\}$.  
$\Omega$ is the energy difference between the bare energy levels
$[|1\rangle\pm|-1\rangle]/\sqrt{2}$ and $\epsilon$ is the coupling
coefficient between the quantum and classical systems.
The variable $x$ satisfies the classical Eq.~(\ref{stocha_norm}). 
We have seen that the spectrum $I(\omega,T)$ of the anharmonic
oscillator is a decreasing function of temperature only for sufficiently small
values of $|\omega|$.
If the quantum system has a dynamics sufficiently slower than the classical
hopping rate ($\Omega\ll R$), then it is sensitive only to the low frequency
components of the bath spectrum. We satisfy the condition $\Omega\ll R$ by 
assuming that the two levels are nearly degenerate. For the sake of simplicity 
we set $\Omega=0$. If $|\psi(0)\rangle=(|-1\rangle+|1\rangle)/\sqrt2$ is
the quantum state at time $t=0$, then at a subsequent time $t$
the density matrix associated with a single realization of $x(t)$ is
$\hat\rho[x](t)=\frac{1}{2}\left(\begin{array}{cc}
1  &  e^{-2i\epsilon\int_0^t x(t)dt}  \\
e^{2i\epsilon\int_0^t x(t)dt} & 1
\end{array}\right)$
and it is a functional of the thermal bath variable. This dependence is removed
by averaging over every realization of $x(t)$. The reduced density matrix 
is $<\hat\rho[x]>_x\equiv\hat\rho_A$  and 
the modulus of its off-diagonal elements 
\be\label{coherence}
|\langle e^{2 i\epsilon\int_0^t x(t)dt} \rangle|\equiv {\cal C}(t).
\ee
is a measure of the coherence between the
superposed states $|-1\rangle$ and $|1\rangle$. 
$\cal C$ is a good indicator since $|-1\rangle$ and $|1\rangle$ are two perfect 
pointer states of a pure dephasing process.
It ranges between the values $0$ (no coherence) and $1$ (complete coherence).

We will show that the decoherence rate of $A$ is proportional
to $I(0,T)$, provided that $\epsilon$ is sufficiently small.
Let $t_c$ be the correlation time of $x(t)$ and $\delta t\gg t_c$ be a
finite time interval. Let us consider the finite difference
${\cal C}(t+\delta t)-{\cal C}(t)={\cal C}(t)
\left[|\langle e^{2 i\epsilon\int_t^{t+\delta t} x(t)dt} \rangle|-1\right]$.
For a large temperature range, we can assume $\langle x^2\rangle\approx1$, thus, 
if $2\epsilon \delta t \ll 1$, then we can approximate the previous equation as 
$
{\cal C}(t+\delta t)-{\cal C}(t)\simeq-2\epsilon^2{\cal C}(t)
\int_t^{t+\delta t}dt'\int_t^{t+\delta t} d\bar t\langle x(t') x(\bar t)\rangle.
$
Since $\delta t\gg t_c$, 
we can replace the second integral with $I(0,T)$ and have
$\dot{\cal C}=-2\epsilon^2 I(0,T)  {\cal C}\equiv -D(T){\cal C}$.
The decoherence rate $D(T)$, being proportional to $I(0,T)$, is a 
decreasing function of $T$ (see Fig.~\ref{fig2}). 
This equation is correct when the integration step $\delta t$ is much smaller
than $1/(2\epsilon)$ and much larger than $t_c$, thus the condition
$1/t_c\gg 2\epsilon$
has to be satisfied. For $\gamma_1=0.4$, the correlation time is about $10$ at 
$T\sim0.25$ and decreases for higher temperatures. At $T\sim1$-$2$, $t_c$ is for 
example about $5$. Thus, $\epsilon$ has to be smaller than about $0.1$-$0.05$.
This classical model demonstrates that the quantum decoherence rate is a 
decreasing function of the temperature when the quantum system is resonant 
with bath spectral components depleted by the thermal energy supplying.
The non-linearity is a fundamental ingredient, since for linear 
baths the spectrum grows linearly with temperature at any frequency.  

Our classical model works when high order
energy levels of the double-well system are activated, i.e., for a 
sufficiently high temperature $T$. Now, we consider the opposite limit
and assume that only the two lowest energy states are populated.
This is possible if the energy difference $\Delta$ between the ground 
and first excited states is much smaller than the higher transition energies.
$\Delta$ is the quantum tunneling frequency between the two wells.
Let the Pauli matrix $\hat\sigma^3_B$ be
the position operator, whose eigenstates $|\pm 1\rangle$ correspond to
the particle in the left or right well. This quantum system ($B$) is 
coupled to a boson bath (spin-boson model) and to the two-state
probe system ($A$) described the Pauli matrices $\hat\sigma^i_A$. 
The overall system $B$-(boson bath) is the non-linear
thermal bath of $A$ (Fig.~\ref{fig3}). 

More precisely, we study a model with the Hamiltonian 
$\hat H=\frac{\Omega}{2} \hat\sigma^1_A+\frac{\Delta}{2}\hat\sigma^1_B+
\epsilon\hat\sigma^3_A\hat\sigma^3_B+ 
\sum_k[\eta\hat\sigma^3_B(\hat a_k+\hat a_k^\dagger)+
\omega_k \hat a_k^\dagger\hat a_k]$,
where $\hat a_k$ and $\hat a_k^\dagger$ are the annihilation and
creation operators of the bath oscillators.
In the present model $x$ is replaced by the Pauli matrix $\hat\sigma^3_B$.
In order to understand how the decoherence of $A$ depends on its bare 
frequency $\Omega$, we have to evaluate the two-time correlation function
$\langle\hat\sigma^3_B(0)\hat\sigma^3_B(t)\rangle$ in the Heisenberg 
representation for $\epsilon=0$, as done with the first model. 
For $\eta$ sufficiently small (see below for a more precise condition), 
we can use the rotating-wave approximation and 
find the master equation for $B$ (section 10.5 of Ref.~\cite{milburn})
\be\label{master}
\frac{\partial\hat\rho_B}{\partial t}=-i\frac{\Delta}{2}[\sigma_B^1,\hat\rho_B]
+{\cal L}\hat\rho_B,
\ee
where
\be\begin{array}{l}
{\cal L}\rho_B\equiv-\gamma_b \bar n\left[\hat\sigma^-_B\hat\sigma^+_B\hat\rho_B
+\hat\rho_B\hat\sigma^-_B\hat\sigma^+_B
-2\hat\sigma^+_B\hat\rho_B\hat\sigma^-_B \right] \\
-\gamma_b (\bar n+1)\left[\hat\sigma^+_B\hat\sigma^-_B\hat\rho_B+
\hat\rho_B\hat\sigma^+_B\hat\sigma^-_B-2\hat\sigma^-_B\hat\rho_B\hat\sigma^+_B
\right]
\end{array}
\ee
and
$\hat\sigma^\pm_B \equiv(\hat\sigma^2_B\pm i\hat\sigma^3_B)/2$.
$\gamma_b$ is a constant coefficient which depends on $\eta$ 
and the density of the boson modes at the frequency $\omega_k=\Delta$ 
and $\bar n(T)\equiv[\exp(\Delta/T)-1]^{-1}$ is their expectation
value $\langle\hat a_k^\dagger\hat a_k\rangle$.
This equation is well-known since it describes the decay of a 
two-level atom coupled to optical modes.

From the master equation~(\ref{master}) and using the quantum regression
theorem, we have that the correlation function at equilibrium 
is~\cite{milburn}
\be\label{corre_B}
\langle\hat\sigma^3_B(0)\hat\sigma^3_B(t)\rangle=
e^{-|t|/\tau_c(T)}\left[e^{-i\Delta t}n_\uparrow+e^{i\Delta t}n_\downarrow\right],
\ee
where $n_\uparrow\equiv\langle\hat\sigma_B^+\hat\sigma_B^-\rangle=\frac{e^{-\Delta/2T}}
{2\cosh(\Delta/2T)}$
and $n_\downarrow\equiv\langle\hat\sigma_B^-\hat\sigma_B^+\rangle=1-n_\uparrow$
are the probability weights for the energy levels of $B$
and  $\tau_c(T)\equiv [(2\bar n+1)\gamma_b]^{-1}$ is the temperature-dependent 
correlation time.
The corresponding spectral function is
$I(\omega,T)=\frac{2\tau_c n_\uparrow}{1+\tau_c^2 (\omega-\Delta)^2}+
\frac{2\tau_c n_\downarrow}{1+\tau_c^2 (\omega+\Delta)^2}$,
which
has two peaks centered in $\pm\Delta$, whose areas
and widths are constant and linear functions of temperature, respectively. 
The rotating-wave
approximation is suitable when the two peaks have a negligible
overlap, i.e., when $\Delta\gg1/\tau_c$.
It is clear that for $\omega\ne\Delta$, the 
spectrum initially increases with temperature, reaches a maximum,
then decreases and tends to zero for $T\rightarrow\infty$. In the
particular case $\omega=\Delta$, $I(\omega,T)$ is a monotonic decreasing
function for any value of temperature. With this in mind, we
choose the frequency $\Omega$ of system $A$ in order to
maximize the effect of temperature on the decoherence reduction.
The system $A$ has to be resonant with the part of the spectrum 
$I(\omega,T)$ which is strongly depleted by the temperature
increase, thus we assume $\Omega=\Delta$. This case simplifies
the analysis and is sufficient for our purposes.

At this point, we take $\epsilon\ne0$ and write out the master
equation for $A$ and $B$. If $\epsilon\ll 1/\tau_c$, then
it is obtained from Eq.~(\ref{master}) by
adding the bare Hamiltonian of $A$ and the interaction
Hamiltonian $\epsilon\hat\sigma_A^3\hat\sigma_B^3$ to
the bare Hamiltonian of $B$.
In the rotating-wave representation (interaction picture), 
the overall bare Hamiltonian 
$(\Delta\hat\sigma^1_A+\Delta\hat\sigma^1_B)/2$
is removed and the following replacements are
performed:
\be\begin{array}{l}
\hat\sigma^3_{A,B}\rightarrow
\hat\sigma^-_{A,B} e^{-i\Delta t}+\hat\sigma^+_{A,B} e^{i\Delta t}.
\end{array}
\ee
The condition $\epsilon\ll1/\tau_c$ justifies a 
rotating-wave approximation also for the interaction between
$A$ and $B$. Thus we obtain the master equation
\be\label{final_master}
\frac{\partial\hat\rho}{\partial t}=
-i\epsilon[\hat\sigma^+_A\hat\sigma^-_B+\hat\sigma^-_A\hat\sigma^+_B
,\hat\rho]+{\cal L}\hat\rho.
\ee
In order to eliminate the parameter $\Delta$ in $\bar n$, we define
the rescaled temperature $\tilde T\equiv T/\Delta$.

\begin{figure} [h!]
\epsfig{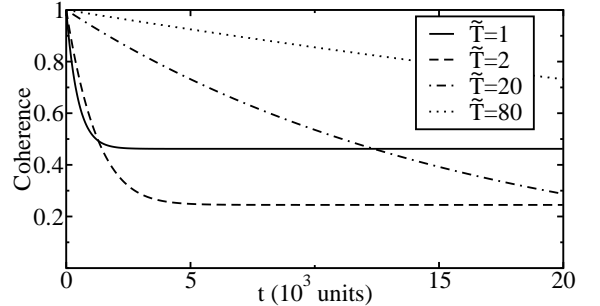}
\caption{Numerically evaluated coherence ${\cal C}$ as a function 
of time for some values of temperature $\tilde T$.}
\label{fig4}
\end{figure}

In this model, the vectors $|1\rangle$ and $|-1\rangle$ are not perfect 
pointer states because of the presence of the bare Hamiltonian of
the system $A$, thus we can not use the previous definition of
coherence. We need a generalized measure of the deviation
from the unitary evolution. A good indicator is provided by 
the quantity $C=\sqrt{2\text{Tr}[\hat\rho_A^2]-1}$, $\hat\rho_A$ being 
the reduced density operator of the system $A$.
For pure states it is equal to $1$ and tends to zero
for completely incoherent high temperature thermal states.
It is easy to prove that this definition is equivalent to
Eq.~(\ref{coherence}) when $|1\rangle$ and $|-1\rangle$ are equally 
populated pointer states. 
At thermal equilibrium, we have ${\cal C}=\tanh[1/(2 \tilde T)]$.
Obviously, it is a decreasing function of temperature and
for $\tilde T\rightarrow0,\infty$ takes
the maximal and minimal values $1$ and $0$, respectively.
The transient behavior of $\cal C$ is numerically evaluated 
by setting the probe system $A$ in the pure state 
$[|-1\rangle+|1\rangle]/\sqrt2$ at the initial time $t=0$, 
with the system $B$ thermalized at the temperature $\tilde T$.
In Fig.~\ref{fig4} we report ${\cal C}(t)$ as a function of time 
for $\gamma_b=1$, $\epsilon=5\cdot10^{-2}$ and some values of 
$\tilde T$. Note that, for $\tilde T=80$, the decoherence rate 
becomes very small with respect to the value at low temperature.

The master equation~(\ref{final_master}) has been obtained
by means of the conditions $\epsilon\ll1/\tau_c\ll\Delta$.
Under the constraint $\epsilon\ll1/\tau_c$, it is possible
to find an analytical relation, since some components of 
$\hat\rho(t)$ can be adiabatically eliminated. The following
approximate solution for the reduced density
operator of the probe system holds,
\be\label{decohe_sb}
\hat\rho_A=\hat\rho_{th}+e^{-\epsilon^2 \tau_c t}
(c_1\hat\sigma^2_A+c_3\hat\sigma^3_A)+e^{-2\epsilon^2\tau_c t} c_2
\hat\sigma^1_A,
\ee
where $\hat\rho_{th}$ is the thermalized density operator,
$c_k$ are constant parameters which depend on the
initial condition and 
$\epsilon^2\tau_c=\epsilon^2\gamma_c^{-1}\tanh(1/2\tilde T)
\equiv\Gamma_d$ 
is a decreasing function of temperature. Notice that
this solution could be obtained directly by the correlation
function~(\ref{corre_B}). By means of the rotating-wave
approximation and the standard hypotheses
used to derive Eq.~(\ref{master}) and the result of
the previous model (second order expansion in the
coupling strength and Markov approximation), one finds that
$\hat\rho_A$ is solution of a master equation identical to
Eq.~(\ref{master}), but with $\gamma_b$ replaced by
$\epsilon^2\tau_c \tanh\frac{1}{2\tilde T}=
\epsilon^2\gamma_c^{-1}\left(\tanh\frac{1}{2\tilde T}\right)^2$. 
Equation~(\ref{decohe_sb}) is its exact solution. This 
derivation will be discussed in an expanded paper.

As in the classical model, we find that the decoherence 
and relaxation rates
are decreasing functions of temperature. In this
case the effect is much stronger. For $\tilde T\gg1$,
the decoherence rate is inversely proportional to the 
temperature and for a sufficiently high thermal noise it 
can be practically reduced to zero. The other interesting feature
is the reduction of $\Gamma_d$ by increasing the coupling
parameter $\gamma_b$. This effects are due to the depletion of 
bath spectrum $I(\omega,T)$ at the frequencies resonant with the
quantum system. 

There is an interesting link between 
the anomalous behaviors in the last model and the Zeno effect. 
Assume that the two systems $A$ and $B$ are in the
state $|-1\rangle_A|1\rangle_B$ at time $t$. Because of their 
mutual coupling, at a subsequent time $t+\Delta t$, they
evolve to the state 
$|-1\rangle_A|1\rangle_B+\epsilon \Delta t |1\rangle_A|-1\rangle_B$
and the probability of the transition 
$|-1\rangle_A|1\rangle_B\rightarrow|1\rangle_A|-1\rangle_B$
is $(\epsilon \Delta t)^2$.
In the meantime, the coherence of this superposition is
destroyed by the thermal bath, which acts as an observer
of the pointer states $|-1\rangle_B$ and $|1\rangle_B$.
Thus, if $\tau_c$ is the correlation time of $B$,
the probability of the transition after a time 
$\Delta t_0\gg \tau_c$ is $P_{(-1,1)\rightarrow (1,-1)}\simeq 
[\epsilon\tau_c]^2\frac{\Delta t_0}{\tau_c}=\epsilon^2\tau_c\Delta t_0$
and we obtain heuristically the relaxation rate in 
Eq.~(\ref{decohe_sb}). Alternatively, the anomalous effect
can be explained as due to a rapid change of the phases of
$|-1\rangle_A|1\rangle_B$ and $|1\rangle_A|-1\rangle_B$
which reduces the coherence of the two states and
consequently their transition probability. 
It is interesting to note that
the decoherence rate suppression for $T\rightarrow\infty$
in the last model can be obtained by a condition established
in Ref.~\cite{pascazio} [equation (25) in that paper]. 
By means of this, one finds that the Hilbert space of 
$A$ is decoherence-free, provided that $\bar n\gg1$.

In conclusion, we have consider two examples of non-linear
baths weakly coupled to a quantum system $A$ and shown
that the decoherence rate of $A$ is a monotonic decreasing
function of temperature. This anomalous behavior is
due to the temperature dependence of the spectral profile in 
non-linear baths.
It is important to realize that at a microscopic scale
non-linear effects can be relevant, as it occurs with
thermal rectifiers. For more realistic models, the 
decoherence rate could not be a monotonic decreasing 
function of temperature. However our analysis indicates
that the decoherence rate is not necessarily minimal
at $T=0$, as for linear baths. Decoherence reduction 
by the thermal energy increase can have relevant
consequences for the design of quantum computers at
nanometer scale. An experimental test of this effect
is feasible with the present technology and
can be observed in nanometer devices~\cite{wallraff,garcia}.
For example, $A$ and $B$ could be a
superconducting qubit and a microwave resonator~\cite{wallraff}.
This work was supported by Ente Cassa di Risparmio di Firenze under the project 
"Dinamiche cerebrali caotiche".

\end{document}